\documentclass[letterpaper,11pt]{article}
\pdfoutput=1
\usepackage{jheppub}
\usepackage{mathtools,amssymb,braket,mathrsfs,tikz,xspace,xcolor,float,color,siunitx,comment,afterpage,multirow,array,hhline,amsthm,framed,placeins,subfig,graphicx}
\usepackage[utf8]{inputenc}
\usepackage[countmax]{subfloat}

\DeclareRobustCommand{\Fig}[1]{Fig.~\ref{#1}}

\DeclareRobustCommand{\Eq}[1]{Eq.~(\ref{#1})}

\DeclareRobustCommand{\Ref}[1]{Ref.~\cite{#1}}
\DeclareRobustCommand{\Refs}[1]{Refs.~\cite{#1}}
\providecommand{\href}[2]{#2}

\title{Going off topics to demix quark \\ and gluon jets in $\alpha_S$ extractions} 

\author[1]{Matt~LeBlanc,}
\author[2]{Benjamin~Nachman,}
\author[3]{and~Christof~Sauer}

\affiliation[1]{Experimental Physics Department, Organisation Europ\'{e}enne pour la Recherche Nucl\'{e}aire (CERN), F-01631 Pr\'evessin Cedex, France -- CH-1211 Genève 23, Geneva, Switzerland}
\affiliation[2]{Physics Division, Lawrence Berkeley National Laboratory,\\ 1 Cyclotron Road, Berkeley, CA 94720, U.S.A.}
\affiliation[3]{Ruprecht-Karls-Universit\"at Heidelberg, Im Neuenheimer Feld 227, Heidelberg, Germany}

\emailAdd{matt.leblanc@cern.ch}
\emailAdd{bpnachman@lbl.gov}
\emailAdd{csauer@physi.uni-heidelberg.de}

\arxivnumber{2206.10642}

\abstract{
Quantum chromodynamics is the theory of the strong interaction between quarks and gluons; the coupling strength of the interaction, $\alpha_S$, is the least precisely-known of all interactions in nature.
An extraction of the strong coupling from the radiation pattern within jets would provide a complementary approach to conventional extractions from jet production rates and hadronic event shapes, and would be a key achievement of jet substructure at the Large Hadron Collider (LHC).
Presently, the relative fraction of quark and gluon jets in a sample is the limiting factor in such extractions, as this fraction is degenerate with the value of $\alpha_S$ for the most well-understood observables. 
To overcome this limitation, we apply recently proposed techniques to statistically demix multiple mixtures of jets and obtain purified quark and gluon distributions based on an operational definiton.
We illustrate that studying quark and gluon jet substructure separately can significantly improve the sensitivity of such extractions of the strong coupling.
We also discuss how using machine learning techniques or infrared- and collinear-unsafe information can improve the demixing performance without the loss of theoretical control.
While theoretical research is required to connect the extract topics with the quark and gluon objects in cross section calculations, our study illustrates the potential of demixing to reduce the dominant uncertainty for the $\alpha_S$ extraction from jet substructure at the LHC.
}

\begin{document}

\flushbottom
\maketitle

\section{Introduction}
\label{sec:intro}

The Higgs field may be the origin of fundamental particle masses, but the strong force is responsible for most of the everyday mass around us.
In the context of Quantum Chromodynamics (QCD), nuclear mass is governed by the strong coupling constant\footnote{The couplings of the Standard Model run with energy; when referring to $\alpha_S$ as a constant, we mean at the scale of the $Z$ boson mass.}, $\alpha_S$.
This fundamental parameter of the Standard Model plays an essential role in particle and nuclear physics, and yet it is the least-precisely known coupling in nature. Aside from the quark masses, the value of $\alpha_S$ is the only free parameter in the QCD Lagrangian.

Given the increasing perturbative precision of many state-of-the-art calculations, such as the Higgs boson production cross section (see \emph{e.g.} Ref.~\cite{Chen:2021isd}), the level of uncertainty on $\alpha_S$ has become increasingly relevant.
Currently, the $\alpha_S$ determination with the smallest uncertainty is obtained from averaging multiple lattice gauge QCD calculations~\cite{flavorLatticeAveragingGroup2019iem,Aoki:2021kgd} combined with measured $B$-hadron mass differences:
$$\alpha_S(M_Z) = 0.1182 \pm 0.0008~\text{(lattice)}.$$
The second most precise category of determinations is from event shape measurements in $e^+e^-$ data, where the most precise extraction is performed using the event thrust and theoretical predictions that include analytic modelling of non-perturbative and hadronization effects~\cite{Abbate:2010xh,ALEPH:2003obs,DELPHI:1996oqw,DELPHI:2004omy,L3:1992nwf,OPAL:2004wof,SLD:1994idb}:
$$\alpha_S(M_Z) = 0.1135\pm 0.0010~\text{(}e^+e^-\text{ thrust)}.$$
These two most precise determinations of $\alpha_s$ are in tension with a significance that exceeds $3\sigma$.

To address this tension within the world average determination of the strong coupling, it is essential to perform new, independent, and complementary extractions of $\alpha_S$ using methods that are sensitive to similar physical effects as the two most precise determinations.
An extraction similar to the $e^+e^-$ determination is possible at the Large Hadron Collider (LHC), albeit with less precision.
However, less precise determinations can still be useful to better understand the tension between determinations in $e^+e^-$ collisions and from lattice QCD.

The 2017 SM Working Group report from the Les Houches Workshop on Physics at TeV Colliders (hereafter, \Ref{Bendavid:2018nar}) explored the prospect of extracting $\alpha_S$ from the internal energy distribution of hadronic jets, or \emph{jet substructure} (JSS), at the LHC.
Like the $e^+e^-$ determinations, a jet substructure-based measurement would be most sensitive to resummation,  not fixed-order effects.
Jets are a principle component of LHC physics~\cite{Salam:2009jx}, and JSS has been well-studied theoretically~\cite{Larkoski:2017jix} and experimentally~\cite{Asquith:2018igt}.
The application of jet grooming strengthens the connection between jets in $e^+e^-$ and $pp$, by suppressing contributions from non-perturbative QCD while also mitigating pileup, improving the jet energy and mass resolution, and providing perturbative calculability.
In particular, the soft-drop / modified mass-drop grooming techniques (hereafter, `soft-drop')~\cite{Larkoski:2014wba,Dasgupta:2013ihk} have recently been used to achieve well-beyond leading logarithm precision~\cite{Frye:2016aiz,Frye:2016okc,Marzani:2017kqd,Marzani:2017mva,Kang:2018jwa,Kang:2018vgn}.
The ATLAS and CMS collaborations at the LHC have recently measured such soft-drop observables~\cite{STDM-2017-04,STDM-2017-33,CMS-SMP-16-010,CMS-SMP-20-010}, demonstrating that the achievable experimental and theoretical precisions are competitive.
Given realistic experimental uncertainties, an extraction of $\alpha_S$ with $\sim$10\% precision from LHC data was found to be possible in~\Ref{Bendavid:2018nar}.

The key difference between extractions from $e^{+}e^{-}$ and $pp$ collisions is that the quark/gluon jet composition of the former dataset is almost purely quarks and the latter is a non-trivial mixture (determined by non-perturbative parton distribution functions, or PDFs).
This introduces a significant challenge for $\alpha_S$ extractions at $pp$ colliders, because changes in the strong coupling constant have the same phenomenology as changes to the composition of colliding partons.
This issue is explored in \Ref{Bendavid:2018nar}, where the degeneracy of the extracted $\alpha_S$ value and the sample's composition of quark and gluon jets is illustrated.
This degeneracy occurs due to the leading behavior of the groomed jet mass distributions being dominated by the product $\alpha_SC_i$, where $C_i\in\{C_F,C_A\}$ is the respective color factor for either quark or gluon jets.
The two approaches identified by \Ref{Bendavid:2018nar} to address this problem are to calculate the quark/gluon fractions (an approach relying on PDFs) or to simultaneously fit $\alpha_S$ with the quark/gluon fractions. 
The first approach suffers from sensitivity to non-negligible PDF uncertainties, while the second requires calculations of joint distributions of groomed observables to high precision.
To quote from the Standard Model Working Group report~\cite{Bendavid:2018nar}: ``Without some kind of conceptual breakthrough, though, we expect that the quark/gluon fraction will be a limiting aspect of $\alpha_S$ extractions from jet substructure at the LHC.''

To overcome this limitation, we apply recently-proposed techniques to statistically demix measured samples of jets in order to obtain purified quark and gluon jet distributions along with the fractional composition of the samples.
A fully data-driven method for estimating the quark/gluon fraction of a sample of jets, known as \emph{jet topics}, was recently introduced in \Ref{Metodiev:2018ftz}.
This approach was extended in \Ref{Komiske:2018vkc}, which argued that jet topics could be used to form a model-independent, operational definition of what is meant by a quark or gluon jet at hadron-level.
The topics procedure relies on identifying regions of phase space enriched with quark- and gluon-initiated jets in order to de-mix their underlying distributions. 
Quark and gluon jets are then defined in aggregate as the maximally-separable categories within two jet samples in data with different quark and gluon jet fractions.
ATLAS has recently made measurements of quark- and gluon-like jet topics~\cite{STDM-2017-16} and topics have also been explored with the CMS Open Data~\cite{Komiske:2022vxg}.  Furthermore, topics have been proposed for similar purposes in relativistic heavy ion collisions~\cite{Brewer:2020och,Ying:2022jvy} and related techniques have been explored in the context of new physics searches~\cite{Dillon:2019cqt,Dillon:2020quc,Alvarez:2019knh}.

In this paper, we propose to use topics to break the degeneracy in the extraction of $\alpha_S$.  In particular, we advocate for using observables that are most effective for computing topics and then marginalizing over these observables to demix the calculable jet mass distribution with the extracted topics.  In this approach, the observables used for constructing topics need not be perturbatively calculable and we will explore various options such as the jet constituent multiplicity and the output of a deep neural network.

This paper is organized as follows.  In Section~\ref{sec:method}, we outline an extension of the original jet topics approach to unbinned observables and illustrate how demixed quark and gluon jet observables can be obtained for distributions which were not amenable to the original approach. 
Demixing such observables requires the simultaneous study of mutually-irreducible distributions, which can be infrared- and collinear-safe without the loss of theoretical control on the distribution of interest.
Then, in Section~\ref{sec:results}, we illustrate the impact that an \emph{in situ} extraction of the quark and gluon jet fractions can have on an extraction of $\alpha_S$ from jet substructure similar to that outlined in~\Ref{Bendavid:2018nar}.


\section{Disentangling quarks and gluons}
\label{sec:method}

\subsection{Breaking the quark and gluon fraction degeneracy}

The main challenge of extracting the strong coupling from jet substructure observables at the LHC is the issue of the sample quark/gluon fraction's degeneracy with the extracted value of $\alpha_S$.
To overcome this problem, multiple samples with different quark fractions can be used.
We consider the two-sample case and call the ``mixtures'' $M_1$ and $M_2$, with probability densities of jet substructure observables $p_{M_1}(\mathcal O)$ and $p_{M_2}(\mathcal O)$.  These densities are statistical mixtures of the underlying quark and gluon jet distributions $p_q(\mathcal O)$ and $p_g(\mathcal O)$:
\begin{equation}\label{eq:qgmixture}
p_{M_1}(\mathcal O) = f_1 \,p_q(\mathcal O) + (1-f_1)\,p_g(\mathcal O),\quad\quad
p_{M_2}(\mathcal O) = f_2 \,p_q(\mathcal O) + (1-f_2)\,p_g(\mathcal O).
\end{equation}
where $f_1$ and $f_2$ are the respective quark fractions of samples 1 and 2, and can each take values from 0-100\%.

The problem is now to obtain the quark and gluon distributions from the mixtures without relying on predictions for $f_1$ or $f_2$.
This question may appear ill-posed without further constraints; for example, what would prevent $f_1=1$ and $f_2=0$ with $p_{q}=p_{M_1}$ and $p_g = p_{M_2}$?
A solution to this problem was developed in \Refs{Metodiev:2018ftz,Komiske:2018vkc} using the technology of \emph{topic modeling}~\cite{10.1145/2133806.2133826}.  This tool de-mixes densities by assuming that there are regions of phase space that is approximately pure in quark or gluon jets.
In particular, the degeneracy is broken by maximally subtracting the mixed distributions from one-another, yielding the distributions of topics $p_{T_1}$ and $p_{T_2}$:
\begin{equation}\label{eq:topics}
p_{T_1}(\mathcal O) = \frac{p_{M_1}(\mathcal O) - \kappa_{12} \,p_{M_2}(\mathcal O)}{1 - \kappa_{12}},\quad\quad
p_{T_2}(\mathcal O) = \frac{p_{M_2}(\mathcal O) - \kappa_{21} \,p_{M_1}(\mathcal O)}{1 - \kappa_{21}},
\end{equation}
where the \emph{reducibility factors} are defined as:
\begin{equation}\label{eq:redfs}
 \kappa_{12} = \inf_{\mathcal O} \frac{p_{M_1}(\mathcal O)}{p_{M_2}(\mathcal O)},\quad\quad \kappa_{21} = \inf_{\mathcal O} \frac{p_{M_2}(\mathcal O)}{p_{M_1}(\mathcal O)}.
\end{equation}

The two topics will correspond to the quark and gluon distributions if quarks and gluons are \emph{mutually irreducible} in the chosen phase space, \emph{i.e.} the spectra have pure regions.
The condition of mutual irreducibility can be expressed as the condition that the quark-gluon reducibility factors are zero, indicating that there is no purer description of the admixtures than by quarks and gluons.
Symbolically, quark-gluon reducibility factors are defined as in \Eq{eq:redfs}:
\begin{equation}\label{eq:qgmi}
\kappa_{qg} = \inf_{\mathcal O}\frac{p_q(\mathcal O)}{p_g(\mathcal O)},\quad\quad
\kappa_{gq} = \inf_{\mathcal O}\frac{p_g(\mathcal O)}{p_q(\mathcal O)}.
\end{equation}

While the groomed jet mass is the most precisely predicted JSS observable at the LHC, it is not mutually irreducible.  In fact, this is a generic property of jet substructure observables that are mostly determined by the hardest emission within a jet (Casimir-scaling).  However, the constituent multiplicity, $n$, in a jet is a suitable candidate for defining a mutually irreducible phase space~\cite{Metodiev:2018ftz,Komiske:2018vkc}.  For any particular multiplicity, the quark-to-gluon likelihood ratio is not zero, but due to the larger color factor for gluons, $p_q(n)/p_g(n)\rightarrow 0$ as $n\rightarrow\infty$.

With the reducibility factors in hand, the quark fractions of the two samples are simply:
\begin{equation}\label{eq:fracs} 
f_1 = \frac{1-\kappa_{12}}{1-\kappa_{12}\kappa_{21}},\quad\quad f_2 = \frac{\kappa_{21}(1-\kappa_{12})}{1-\kappa_{12}\kappa_{21}}\,.
\end{equation}
Thus, obtaining the reducibility factors by studying pure substructure phase space regions provides a clean way to extract the quark/gluon jet composition of the samples and circumvent the degeneracy between this quantity and the strong coupling.
Note that once we have $f_1$ and $f_2$, we can ignore the observable(s) used to compute them since the quark/gluon fraction is a property of an event ensemble, not a property of an observable ensemble.
This means that the fractions could be computed with arbitrarily complex observables while the $\alpha_s$ extraction could be performed with precisely predicted observables like the groomed jet mass.
We will use this flexibility to use infrared and collinear unsafe observables (like constituent multiplicity) as well as neural networks.
Multiple observables can be used to estimate the systematic uncertainty due to the non-mutual-irreduciblity of the selected observables.
If all observables are mutually irreducible, then they should all result in the same extracted quark/gluon fractions.

In practice, a procedure to obtain the reducibility factors with finite statistics is required in place of the infimum in \Eq{eq:redfs}.
\Refs{Metodiev:2018ftz,Komiske:2018vkc} describe a binned procedure to compute the reducibility factors.
We use specified anchor phase space regions that compromise purity and statistical precision\footnote{While our paper was in the final stages of completion, Ref.~\cite{Komiske:2022vxg} introduced another approach.
We leave the exploration of different approaches and the impact on an $\alpha_s$ extraction to future studies.
See also Ref.~\cite{Brewer:2020och} for another fraction finding procedure.}.
These anchor phase space regions $\Phi_q$ and $\Phi_g$ are specified to be highly pure in quark and gluon jets, respectively.
For instance, jets with constituent multiplicities below some low value can be declared to be the anchor region of quark jets, and jets with constituent multiplicities above some high value can be declared to be the anchor region for gluon jets.
Determining the most experimentally robust anchor phase space regions for quarks, gluons, and other types of jets is an interesting direction for future studies.
For example, double-$b$ tagging jets can be an effective way to isolate pure-gluon anchor regions.

With the phase space regions specified and taking $M_2$ to be the more quark-enriched sample, one then calculates the probability mass in the anchor regions in both samples and takes the appropriate ratios:
\begin{equation}\label{eq:redfs22}
\kappa_{12} = \frac{\int d\mathcal O\, p_{M_1}(\mathcal O) \, \Theta(\mathcal O \in \Phi_q)}{\int d\mathcal O\, p_{M_2}(\mathcal O) \, \Theta(\mathcal O \in \Phi_q)},
\quad\quad \kappa_{21} = \frac{\int d\mathcal O\, p_{M_2}(\mathcal O) \, \Theta(\mathcal O \in \Phi_g)}{\int d\mathcal O\, p_{M_1}(\mathcal O) \, \Theta(\mathcal O \in \Phi_g)},
\end{equation}
In the limit where $\Phi_q$ and $\Phi_g$ approach the values of $\mathcal O$ minimizing \Eq{eq:redfs}, the two definitions of the reducibility factors become equivalent.
Hence using \Eq{eq:redfs22} can allow for additional statistics at the cost of a more aggressive definition.
“In this work, anchor regions were selected that result in robust performance across a range of quark- and gluon-initiated jet mixtures, but that do not necessarily result in optimal performance for a specific set of mixtures.
The performance of the method given this choice is discussed further in Sec.~\ref{sec:topicresults}.”
With the reducibility factors calculated in this way, one can then obtain the topics via \Eq{eq:topics} as before.

\FloatBarrier
\section{Probing the strong coupling}
\label{sec:results}

\begin{figure}
    \centering
    \includegraphics[width=0.99\textwidth]{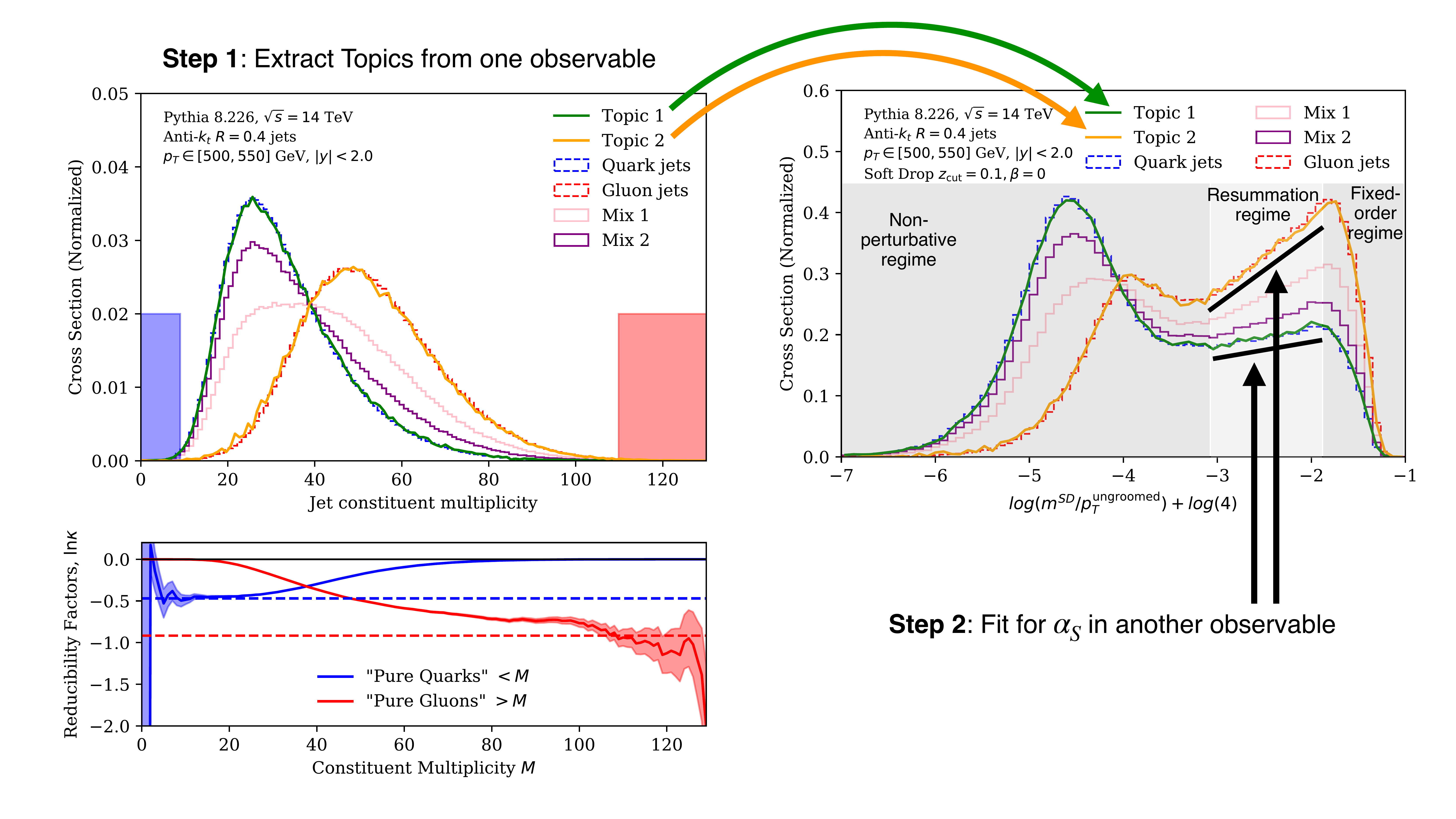}
    \caption{An overview of the proposed extraction of $\alpha_S$. First (left), the topics are extracted from an approximately mutually irreducible observable.  Illustrated above is the case of constituent multiplicity.  Two samples with different quark/gluon compositions are used to extract the topics.  These topics are interpreted as quarks and gluons and the fractions of the two topics are then used as the quark and gluon fractions.  Next (right), an observable with a precisely known cross section is used to fit for $\alpha_S$.  We consider the groomed jet mass and focus on the resummation regime at intermediate mass where non-perturbative and fixed-order effects are small.  The predictions as a function of $\alpha_S$ are then fit to the data, with the proper mixture of the quark and gluon components. }
    \label{fig:overview}
\end{figure}

In this section, we perform a numerical study to demonstrate the extraction of $\alpha_S$ using input from jet topics.  This process is simplified from a full experimental and theoretical analysis in order to concisely illustrate the salient components.  The entire process is summarized in Fig.~\ref{fig:overview}.  First, two samples are used to extract the jet topics using a nearly mutually irreducible observable.  We will study multiple observables, including the constituent multiplicity and a neural network.  For this part, we will use Parton Shower Monte Carlo (PSMC) simulations that describe a complete final state in terms of hadrons.  Next, the inferred topics are interpreted as quark and gluon jets and the corresponding fractions are assigned to the event sample as a whole.  We use the groomed jet mass to perform the actual extraction of $\alpha_S$.  To do this, we use pseudodata generated from resummed calculations, mixing the quark and gluon jet predictions.  These pseudodata are then fit to the calculations with $\alpha_s$ and the quark/gluon fractions left unknown.  While the PSMC predictions qualitatively agree with calculations with leading logarithm accuracy, we use the pseudodata in order to consider more precise predictions and to ensure that the method closes for this part of the analysis.  Systematic uncertainties related to method non-closure will be an important part of future studies on the path towards a full experimental measurement.

\subsection{Extracting the Topics}
\label{sec:extract}

\subsubsection{Event Generation}
\label{sec:evgen}

Quark and gluon jets from proton--proton collisions are generated using \textsc{Pythia} 8.226~\cite{Sjostrand:2014zea} via the $Z(\to \nu\bar\nu)+(u,d,s,c,b)$ and $Z(\to\nu\bar\nu)+g$ processes at $\sqrt{s}=14$ TeV using the default Monash set of tuned parameters~\cite{Skands:2014pea}.
Hadronization and multi-parton interactions (MPI) are included.
Stable final-state particles (non-neutrinos) are clustered into $R=0.4$ anti-$k_t$ jets~\cite{Cacciari:2008gp} using \textsc{FastJet} 3.3.0~\cite{Cacciari:2011ma}. These events are publicly available as a part of the \href{https://energyflow.network}{\texttt{EnergyFlow}} package~\cite{energyflow,qgjets_pythia_2019}.

The leading jet in each event is selected for inclusion in the study if it has transverse momentum $p_T\in[500,550]$ GeV and rapidity $|y|<2.0$.
Two million quark and gluon jets which pass this selection are generated with equal proportions, allowing for the construction of two independent mixed samples with arbitrary $q$/$g$ fractions containing up to 500k jets each.

The generated quark and gluon jets are thus mixed into sub-samples of 500k jets in order to explore the performance of the jet topics technique as a function of the sample composition.
In practice, any two samples with different compositions can be used, such as jets from $Z$+jet and dijet processes, or the more-central and more-forward jets in dijet systems~\cite{Gallicchio:2011xc,STDM-2014-17,STDM-2015-12,STDM-2017-16,STDM-2017-33}.
The choice of samples can, in principle, affect the extracted topics, due to non-universality of the underlying distributions. This can be quantified by considering several different samples, and can be mitigated with jet grooming techniques~\cite{Ellis:2009su,Ellis:2009me,Krohn:2009th,Dasgupta:2013ihk,Larkoski:2014wba,Frye:2016okc,Frye:2016aiz,Marzani:2017mva,Marzani:2017kqd}.
Given the existing theoretical and experimental handles with which to quantify and mitigate process-dependence of the extracted topics, this question is neglected in these studies to instead focus on the performance and precision of the jet topics procedure in extracting the fractions of quark and gluon jets.  The process-dependence is also likely to be sub-dominant to other sources of uncertainty~\cite{1810.05653}.

We emphasize that the \textsc{Pythia}-labeled ``quark'' and ``gluon'' distributions are fundamentally ill-defined, both theoretically and experimentally.
While we use these \textsc{Pythia} quark and gluon jet samples in this study to have a sense of comparison for this approach, we note there is a fundamental limitation to relying on the quark/gluon labels in Monte Carlo generators: only an operational definition of quark and gluon jets as in \Ref{Komiske:2018vkc} is a well-defined object at hadron colliders.
Systematic differences between the two quark/gluon definitions are not necessarily physically meaningful.
For the purpose of extracting $\alpha_S$, what matters most is that the notion of quark and gluon jet used to determine the fractions aligns with the theory definition used in the calculations.  We leave further explorations of this connection to future work.

\subsubsection{Observables}

Quark and gluon jet fractions are extracted in this work from topics obtained using one of two observables: either the jet constituent multiplicity $n_{\text{const}}$, or the output of a particle flow network (PFN) trained to classify quark and gluon jets~\cite{Komiske:2018cqr}.

The PFN is a particle-level network architecture based on the deep sets approach~\cite{DBLP:journals/corr/ZaheerKRPSS17} that is designed to take advantage of the natural structure of jets as unordered, variable-length sets of particle.
Our model is implemented using the EnergyFlow package~\cite{energyflow} with settings that closely follow the PFN implementation in~\Ref{Komiske:2018vkc}, which studied the same samples of simulated quark and gluon jets~\cite{qgjets_pythia_2019}.
It is trained to classify quark and gluon jets according to the unphysical \textsc{Pythia} label, and is provided with lists of the momentum fraction, translated rapidity and translated azimuthal angle of each jet constituent.
The rapidity and azimuthal angle are translated to the origin according to the $E$-scheme jet axis.
The network is implemented and trained in Keras~\cite{keras} with the TensorFlow backend~\cite{tensorflow}. 
Training, validation and testing datasets were constructed from the mixed sample of quark and gluon jets using 1M, 200k and 200k events, respectively.
The model was trained using a batch size of 5000 jets and patience parameter of 10 epochs, monitoring the validation loss.

Other choices of observables were studied, incluing the soft-drop multiplicity~\cite{Frye:2017yrw} with $z_{\text{cut}}=0.1$ and $\beta=0,1,2$ and the output of an energy-flow network (EFN) quark/gluon classifier with similar settings to the PFN, but the results were not found to significantly outperform those shown in Sec.~\ref{sec:topicresults}.

\begin{figure}[htp]
    \centering
    \subfloat[]{\includegraphics[width=0.49\textwidth]{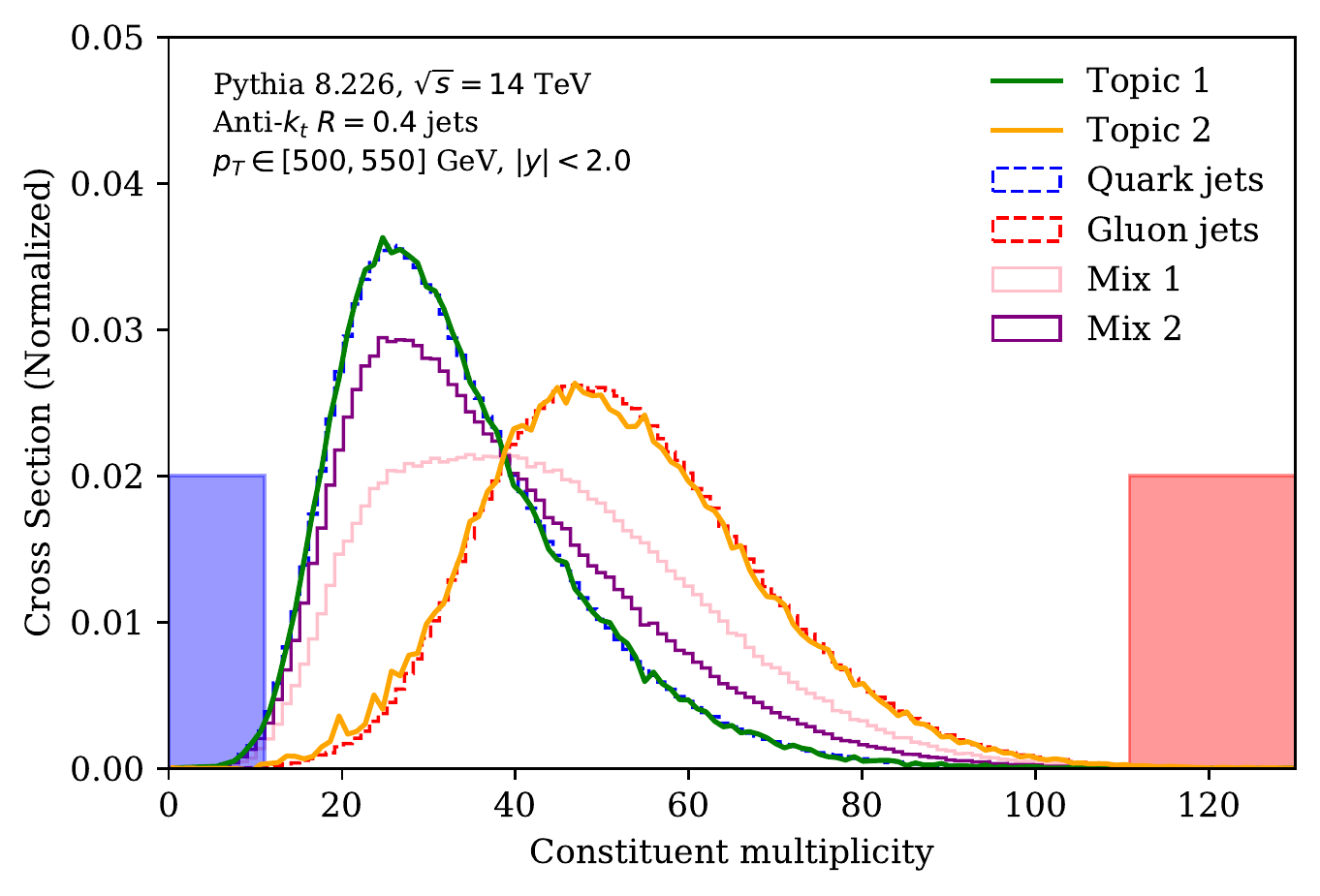}}
    \subfloat[]{\includegraphics[width=0.49\textwidth]{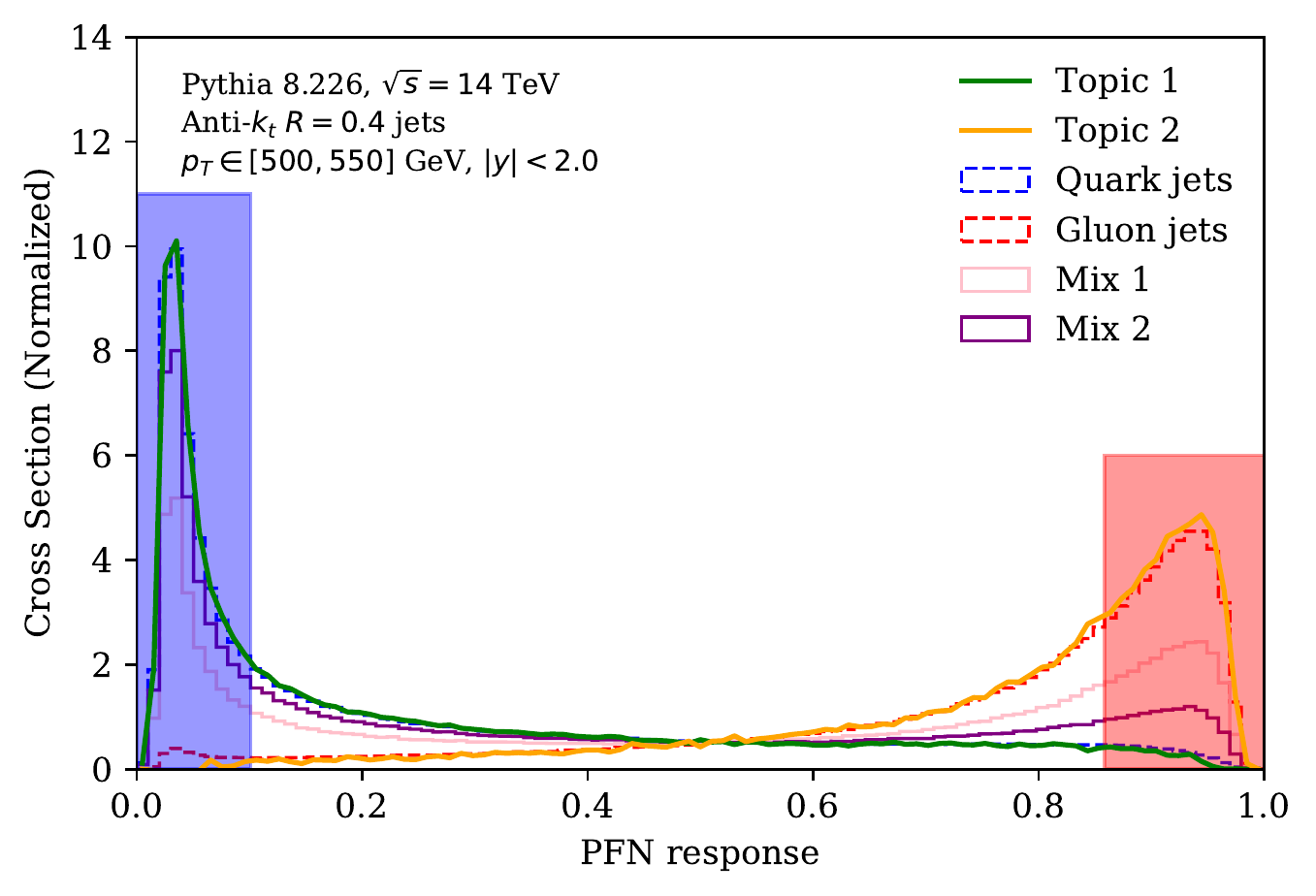}}
    \caption{The distributions of (a) the constituent multiplicity distribution, and (b) the output of a particle flow network trained to classify quark and gluon jets are shown for the mixed samples with 80\% (pink) and 50\% (purple) quark jets, with the pure quark (blue) and gluon (red) distributions from \text{Pythia} and the extracted jet topics (green and orange). 
    The anchor regions used to define quark- and gluon-enriched samples for the topics extraction are indicated by blue and red rectangles at the left and right side of the figures.}
    \label{fig:topics}
\end{figure}

Figure~\ref{fig:topics} illustrates the topics obtained using irreducibility factors from the $n_{\text{const}}$ and PFN distributions for the samples described in Section~\ref{sec:evgen}.
For this demonstration, mixed samples of quark- and gluon-initiated jets that are 80\% and 50\% quark-initiated are used.
There is good agreement between the extracted topics and those obtained from \text{Pythia}, as has been observed in previous studies of jet topic modelling using these observables~\cite{Komiske:2018vkc}.

\subsubsection{Numerical Results}
\label{sec:topicresults}

\begin{figure}[htp]
\centering
\subfloat[True fractions]{\includegraphics[width=0.89\columnwidth]{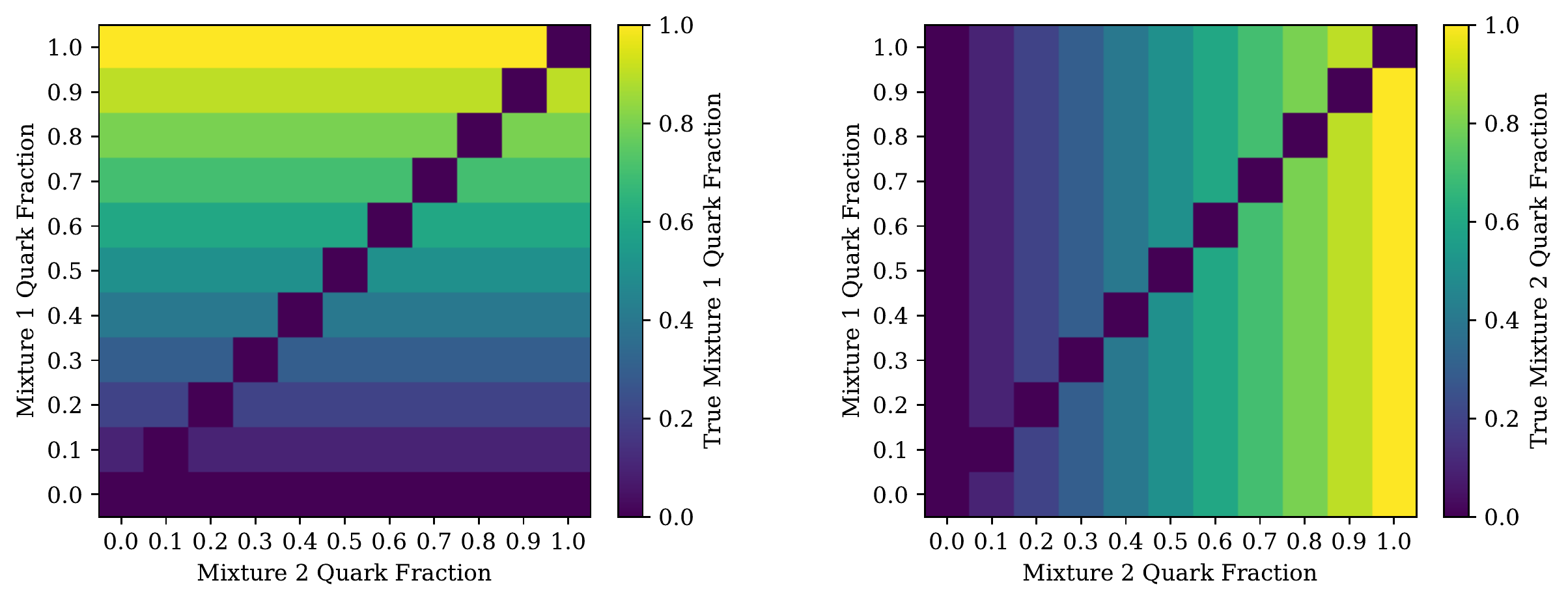}}\\
\subfloat[Constituent multiplicity]{\includegraphics[width=0.89\columnwidth]{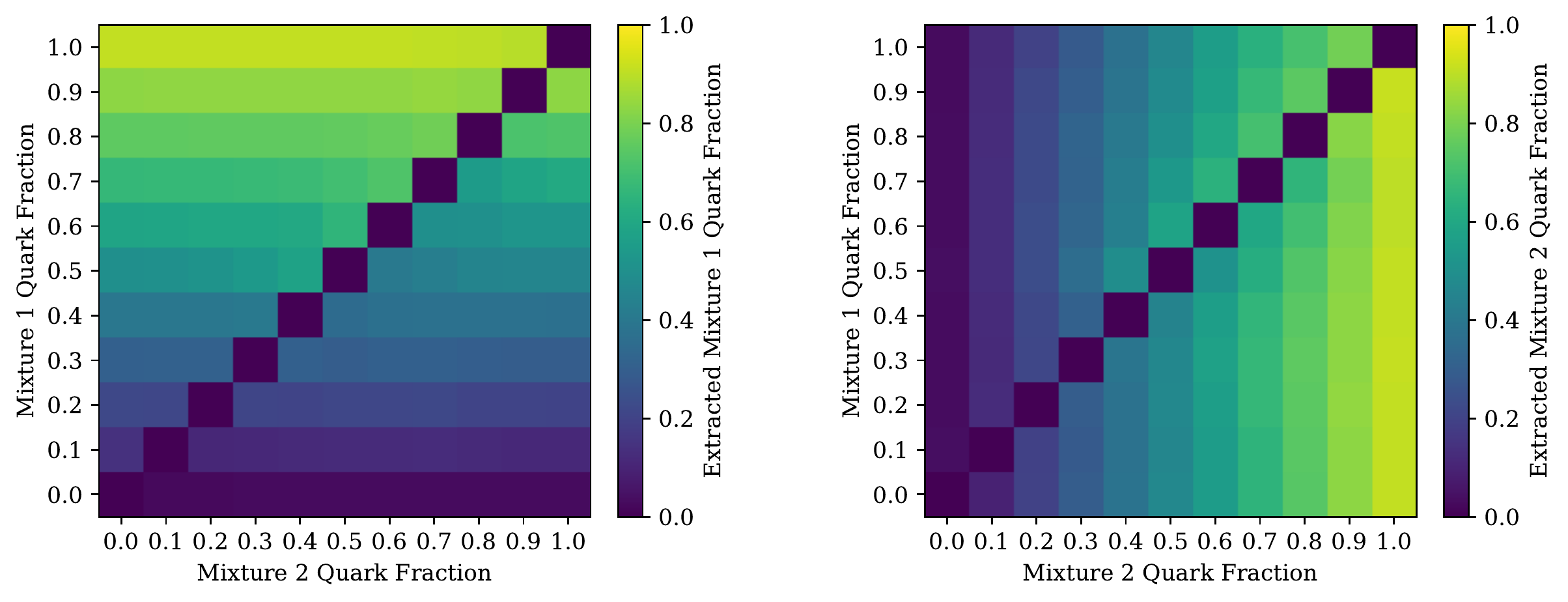}}\\
\subfloat[Particle flow network]{\includegraphics[width=0.89\columnwidth]{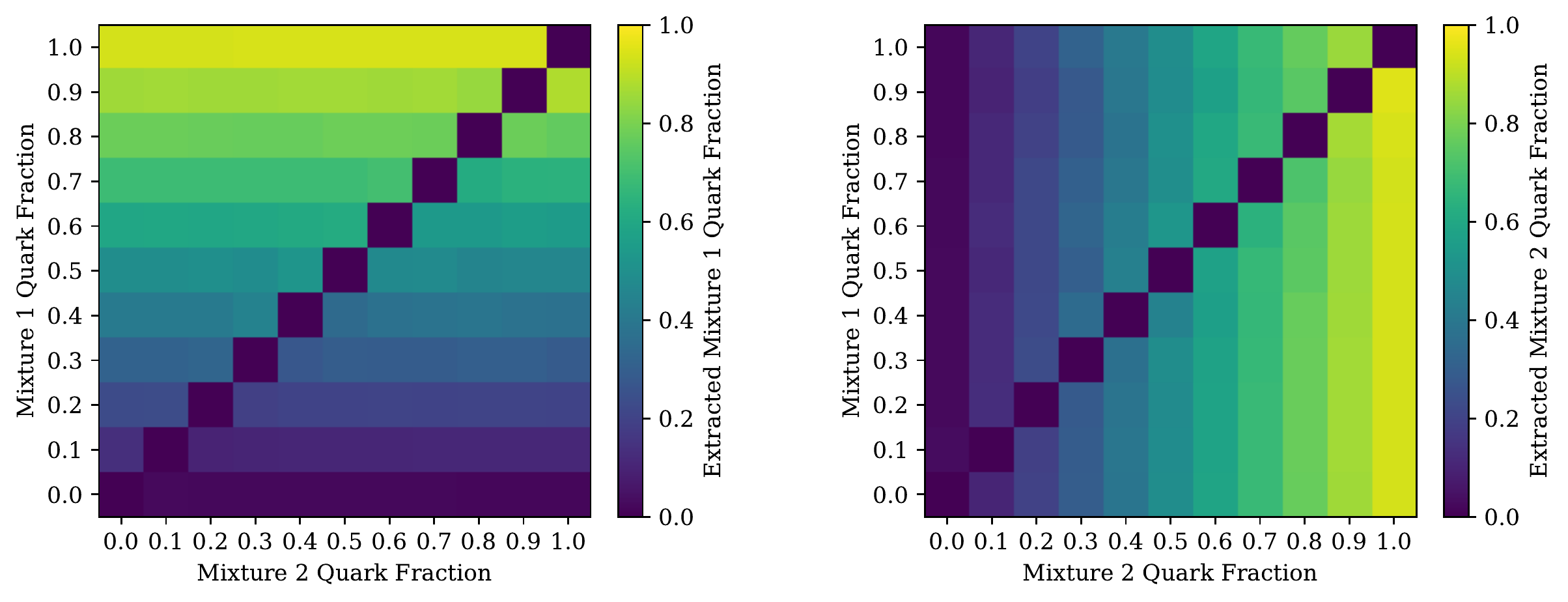}}
\caption{\label{fig:fracs}The (top row) true and (bottom row) extracted quark fractions of (left column) mixture 1 and (right column) mixture 2 for different quark fraction values between zero and one. The (middle row) jet constituent multiplicity or (bottom row) the output of a particle-flow network trained to discriminate quark and gluon jets was used to construct the topics used for the extraction of the quark fractions.}
\end{figure}

The efficacy of the topics-based extraction of the quark and gluon jet fractions can be studied as a function of the quark fractions $f_1$ and $f_2$ of the two mixtures.
We consider all possible mixtures with $f_1,f_2\in[0,1]$ in increments of $0.1$.
The extracted fractions are shown in \Fig{fig:fracs} and compared to the true proportions of the mixtures used for each of the two mixtures.
The topics procedure clearly captures the quark fraction dependence of the samples, with the exception of the equal-fraction case where the problem becomes ill-posed and degenerate.
For these studies, topics were extracted using either the jet constituent multiplicity or the output of a particle-flow network trained to discriminate quark and gluon jets.
Either choice results in an accurate extraction of the quark/gluon jet composition of the underlying sample, but the extracted fractions via PFN tend to exhibit smaller non-closure than those extracted using the particle multiplicity.
A single choice of anchor regions was selected for all extracted fractions (selected for a combination of mixed samples with 50\% and 80\% quark jets), which can cause the performance to vary as a function of the sample composition.
In particular, non-closure can be large when the quark and gluon jet fractions of the two samples are similar.
Such non-closures can result in, for example, quark-initiated jet fractions that are similarly biased (consistently over- or under-estimated) for both mixtures in this region.
This non-closure could be reduced by optimising the anchor regions based on the expected quark-initiated jet purity of the mixed samples used in an analysis, or by following the proposed methodology of Ref.~\cite{Komiske:2022vxg}, which was published concurrently with this work.

\subsection{Fitting the strong coupling}

A detailed analysis of the sensitivity and robustness of various jet substructure observables for extracting the strong coupling from the groomed jet mass was performed in \Ref{Bendavid:2018nar}.
Rather than repeating the full analysis, we directly focus on the case of the groomed jet mass (two-point correlator, $e_2^{(2)}$) with soft-drop parameters $z_{\text{cut}}=0.1$ and $\beta=0$.
The same procedure can be applied to any choice of observable, and so these conclusions apply more broadly.
In these fits, two mixed samples with quark fractions of $50\%$ and $80\%$ are used, corresponding roughly to the respective quark/gluon compositions of dijet and $Z$+jet events at the LHC.

We do not address the question of systematic uncertainties on the measured distribution, as our goal is to highlight the increased sensitivity of the fit in general following this modified approach.

\subsubsection{Analytic Predictions}

We use the same analytic calculations as in \Ref{Bendavid:2018nar}, which are based on formulae from \Ref{Marzani:2017kqd,Marzani:2017mva} at leading logarthim (LL) and next-to-leading logarithm (NLL) resummation.  At lowest order, resummed the normalized differential cross section is given by

\begin{align}
    \frac{1}{\sigma}\frac{d\sigma}{de_2^{(2)}}=-\frac{1}{e_2^{(2)}}\frac{\alpha_S C_i}{\pi}[\log(z_\text{cut}-B_i)]\exp\left[-\frac{\alpha_SC_i}{\pi}[\log(z_\text{cut})-B_i]\log(e_2^{(2)})\right]\,,
\end{align}
where $C_i$ is the color factor that is $C_F=4/3$ for quarks and $C_A=3$ for gluons.  Additionally, $B_i=-3/4$ for quarks and $-\frac{11}{12}+\frac{n_f}{6C_A}$ for gluons (for $n_f$ active quark flavors).  On a log plot, the cross section is approximately linear in the mass with a slope set by $\alpha_S C_i$.  As this coefficient is a product of $\alpha_S$ and the color factor, it is generally not possible to extract both at the same time using the jet mass.

We consider the regime where resummation is expected to describe most of the cross-section:

\begin{align}
0.001\lesssim m^{SD}/p_\mathrm{T}^{\text{ungroomed}}\lesssim0.01\,.
\end{align}
From the calculations, templates are constructed to estimate $p_{q}(\alpha_S)$ and $p_{g}(\alpha_S)$.  Matching to fixed-order predictions is relatively straightforward (at next-to-leading order at least), but computationally expensive, since we do not have an analytic prediction that can be quickly varied as a function of $\alpha_S$.  A complete experimental result would need to include both a matching to fixed order as well as a sensitivity study to non-perturbative effects at low mass.

\begin{figure}[htp]
\centering
\subfloat[]{\includegraphics[width=0.49\columnwidth]{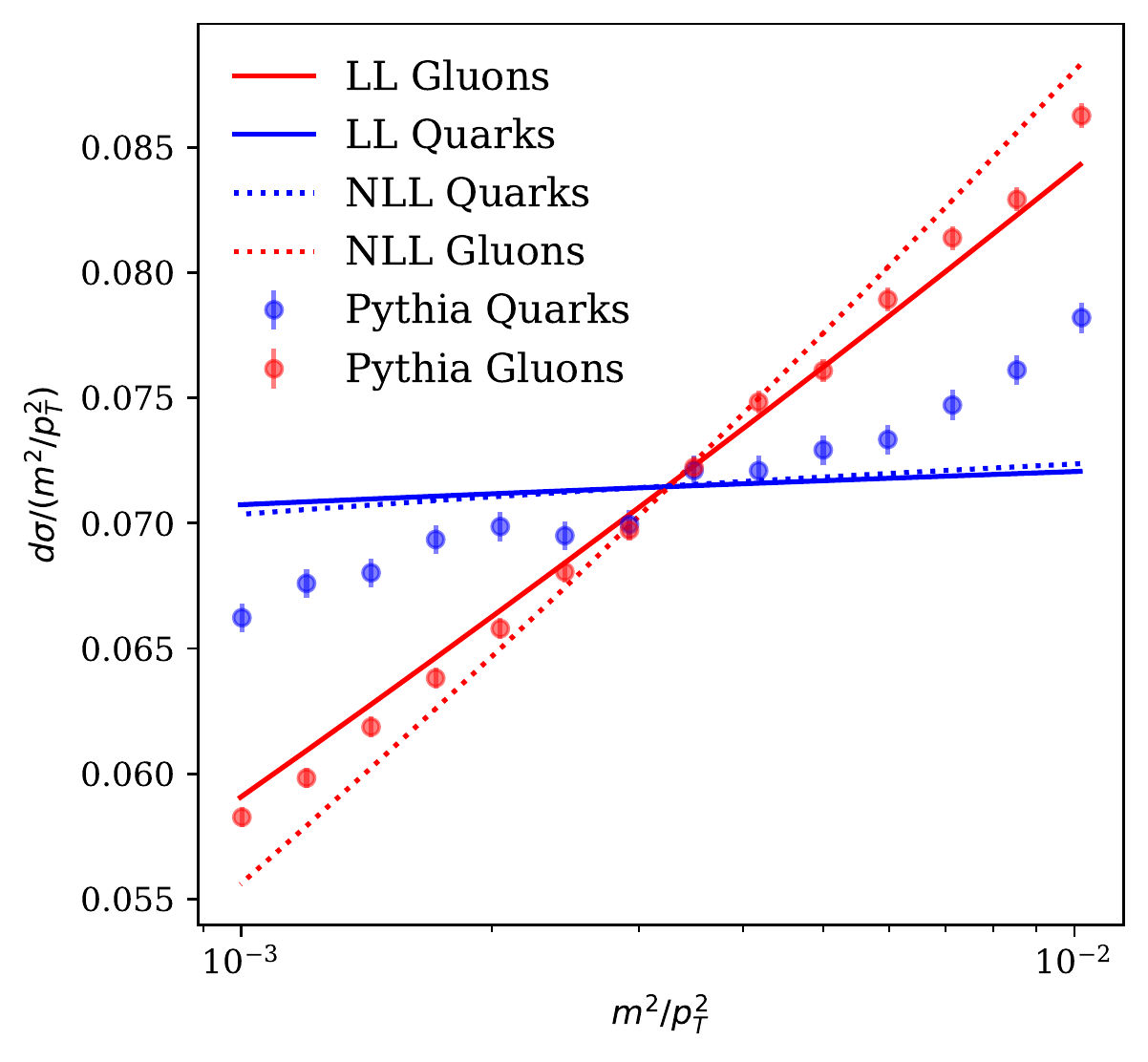}\label{fig:pythia_LL_NLL:a}}
\subfloat[]{\includegraphics[width=0.49\columnwidth]{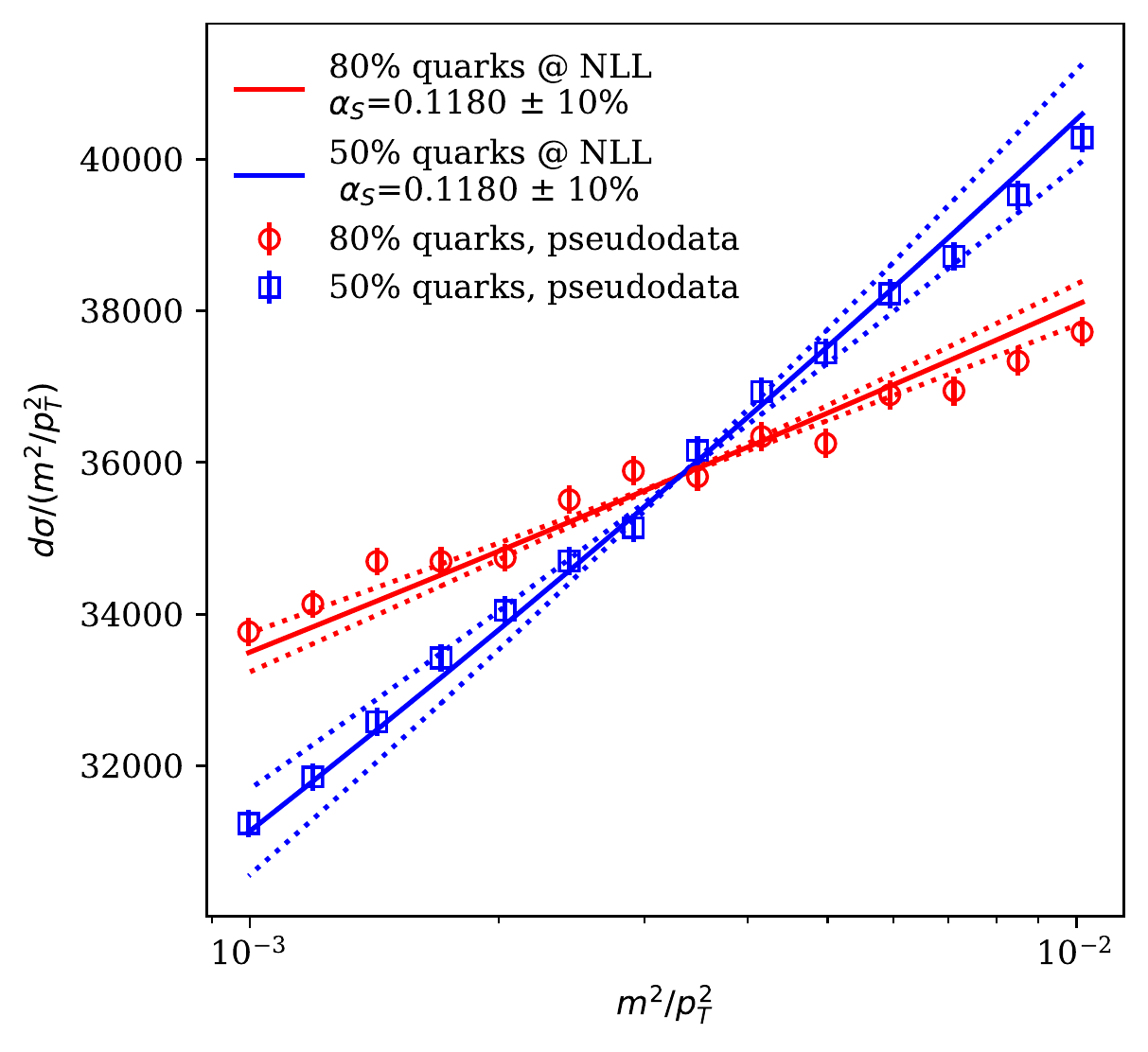}\label{fig:pythia_LL_NLL:b}}
\caption{\label{fig:pythia_LL_NLL} (left) Comparison of Pythia soft-drop mass prediction to LL and NLL theory predictions. (right) Example pseudodata generated from the NLL theory predictions.}
\end{figure}

The NLL predictions for quarks and gluons are compared to the LL expressions and the distributions obtained from \textsc{Pythia} (Sec.~\ref{sec:evgen}) in figure~\ref{fig:pythia_LL_NLL:a}.  There is qualitative agreement between \textsc{Pythia} and LL, with a closer match for gluons than for quarks.  The NLL corrections are negligible for quarks and about 10\% for gluons.  Due to the larger color factor, the dependence on $\alpha_S$ is larger for gluons than for quarks; accordingly, we expect that the statistical precision will be best for a sample of jets enriched in gluon jets.

\subsubsection{Fitting Procedure and Results}

We perform a binned $\chi^2$-type fit using templates generated from the (N)LL predictions.  Example templates for quark-jet fractions $f_1 = 0.5$ and $f_2=0.8$, as well as $\alpha_S=0.1180\pm 10\%$ are shown in Fig.~\ref{fig:pythia_LL_NLL:b}.

The fit itself is implemented using the \verb|scipy.curve_fit| function~\cite{2020SciPy-NMeth};
\begin{equation}
\alpha_S = \text{argmin}_{\alpha_S} \sum_i \frac{(h_i(\alpha_S) - p_i(\alpha_S))^2}{\sigma(h_i(\alpha_S))^2},
\end{equation}
where $p_i$ and $h_i$ are the respective number of pseudodata events or predicted counts from the theoretical prediction in each histogram bin, and $\sigma(h_i)$ is the statistical uncertainty of the theoretical prediction in that bin. This approach closely follows the methodology of \Ref{Bendavid:2018nar}.

\begin{figure}[htp]
\centering
\subfloat[]{\includegraphics[width=0.49\columnwidth]{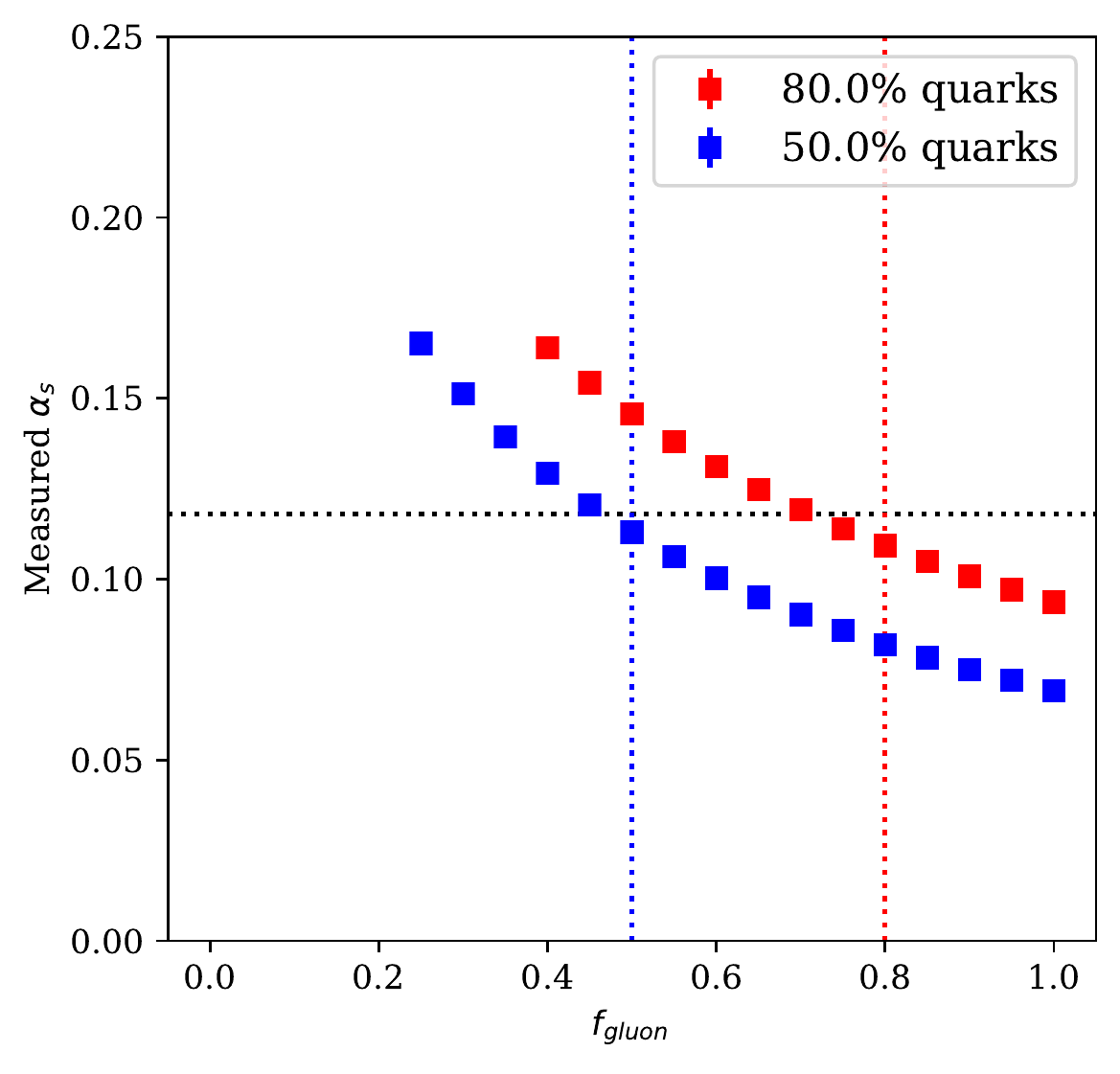}}
\subfloat[]{\includegraphics[width=0.49\columnwidth]{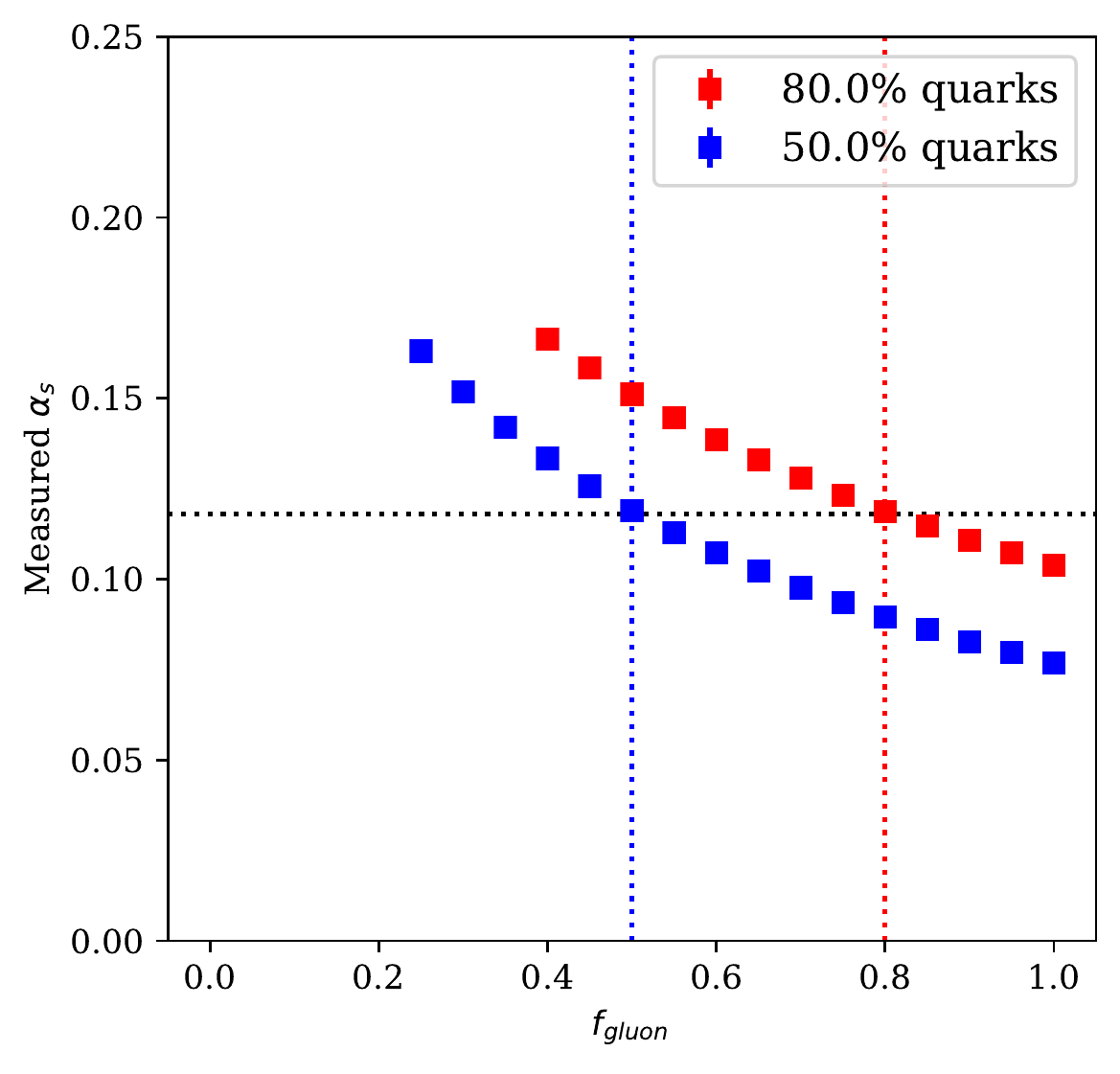}}
\caption{\label{fig:exfit} Characteristic curves obtained by extracting $\alpha_S$ from the groomed jet mass and varying the underlying sample quark/gluon composition using (a) LL and (b) NLL theory predictions.}
\end{figure}

The degeneracy of the extracted $\alpha_S$ and the sample quark/gluon jet composition is illustrated in Fig.~\ref{fig:exfit}, using fits to both the LL and NLL theory predictions.
Each extracted $\alpha_S$ value is obtained by fitting the mixed samples of quark and gluon jets, while specifying the quark and gluon jet fractions to have another value in the fit.
A characteristic curve is obtained from each mixed sample, by extracting the value of $\alpha_S$ while varying the specified fraction of quark jets.

These samples are then fit for the value of $\alpha_S$ and their respective quark and gluon jet fractions simultaneously, following the `Les Houches' methodology of~\Ref{Bendavid:2018nar}.
Such fits are poorly constrained, particularly in terms of the quark/gluon jet fractions, and so large uncertainties and instabilities result from the fitting procedure (particularly when fitting the LL theory prediction)).
A combined fit to both mixed samples to extract their respective quark/gluon fractions and a shared $\alpha_S$ value can result in improved fit performance, although it ultimately suffers from the same shortcomings.

In our topics-based approach, we do not let the quark/gluon fractions float in the fit and instead constrain them to be equal to the fractions that are extracted from the mixed samples using quark and gluon topics (Sec.~\ref{sec:extract}).

Topics extractions from both the jet constituent multiplicity and the output of a PFN trained to classify quark and gluon jets were studied, and provide comparable levels of performance.
The results of this fitting approach are significantly more stable and precise than the original procedure described in \Ref{Bendavid:2018nar}, as only the value of $\alpha_S$ remains unconstrained in the fit.

\begin{figure}[htp]
\centering
\includegraphics[width=0.80\columnwidth]{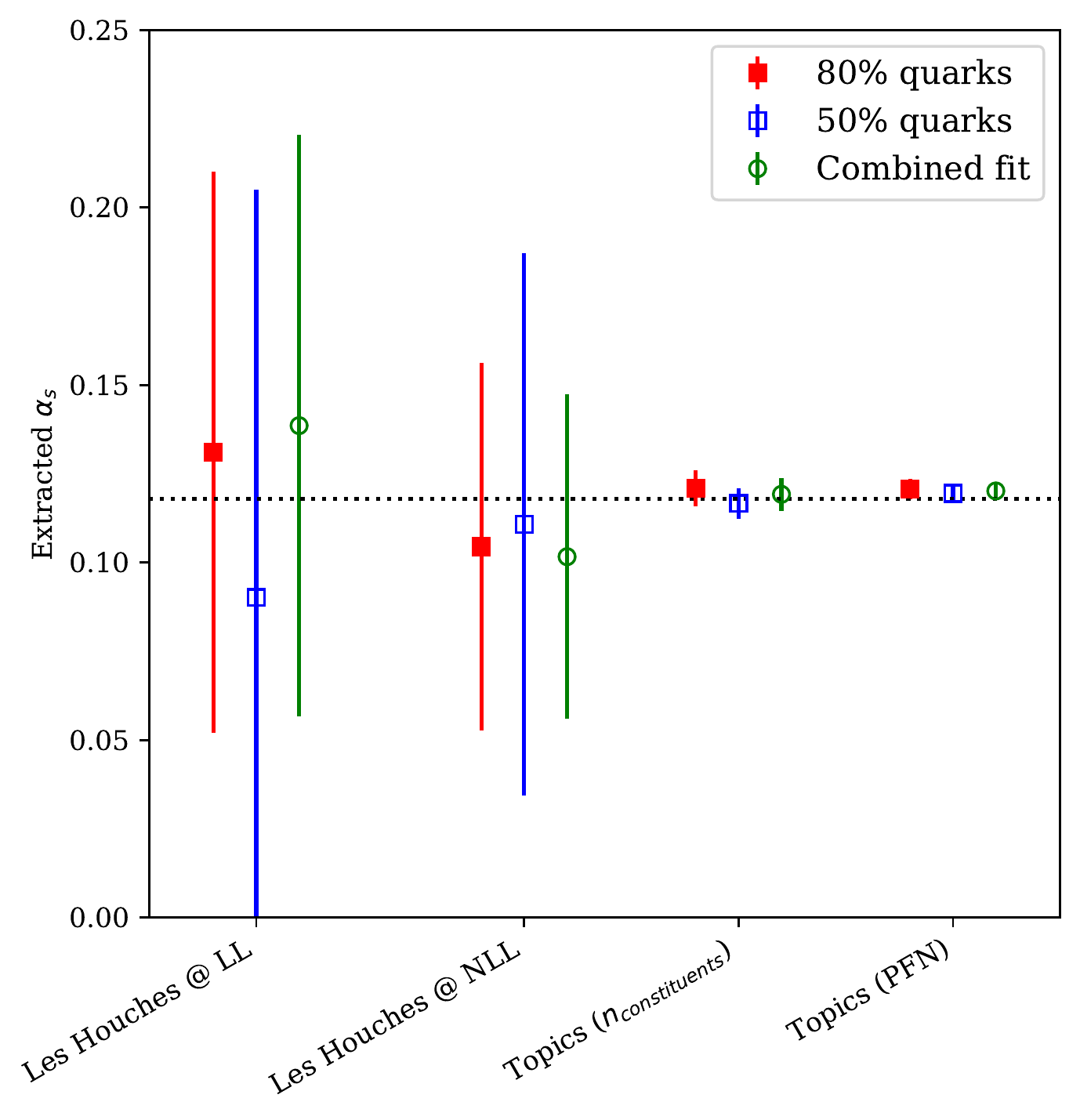}
\caption{\label{fig:ascomp} Value of $\alpha_S$ obtained following the methodology of~\Ref{Bendavid:2018nar} with theoretical predictions at leading-logarithmic and next-to-leading-logarithmic accuracy and no \emph{a priori} knowledge of the quark/gluon fractions, compared to the value obtained using next-to-logarithmic theory and flavor composition information obtained with the jet topics procedure.
}
\end{figure}

A comparison between the extracted values of $\alpha_S$ from the fits described above is provided in figure~\ref{fig:ascomp}.
The extractions labeled as `Les Houches' follow the procedure in \Ref{Bendavid:2018nar} and simultaneously fit for $\alpha_S$ and the quark/gluon fractions.
This results in large uncertainties from the fitting procedure due to the degeneracy between the fitted parameters.
The NLL extraction is slightly more precise than the LL extraction, due to the non-linearity that helps to break the degeneracy between $\alpha_S$ and the quark/gluon fraction in the differential cross section.
As noted in the previous section, the gluon spectra are more sensitive to $\alpha_S$ than the quark distributions.

By using quark/gluon fractions determined from topics, we completely eliminate the uncertainty induced from the degeneracy between $\alpha_S$ and the quark/gluon fraction.
The residual uncertainties are statistical, or related to biases from the non-irreducibility of the observable used to extract the topics.
While there are other theoretical and experimental challenges associated with the extraction of $\alpha_S$ from jet substructure, the topics-based approach opens a way forward to this goal that was previously blocked.

\FloatBarrier
\section{Conclusions}
\label{sec:conclusion}

In this paper, we have outlined a proposal to apply statistical demixing procedures in the context of $\alpha_S$ extractions from groomed jet substructure observables.
This approach breaks a degeneracy between the quark/gluon jet composition of the data sample and the extracted value of the strong coupling.  This significantly reduces the largest source of uncertainty to a manageable level.

The degeneracy is broken by extracting the quark/gluon fraction with one observable and then applying it to another that can be precisely predicted.
An interesting feature of this approach is that the first observable need not have a cross section that is analytically tractable.
In particular, we can use infrared and collinear unsafe and/or machine-learning based observables for precision jet physics.
Future work will explore the relationship between the extracted topics and the partonic degrees of freedom in the precision calculations used to extract $\alpha_S$.

While we have proposed a method to essentially eliminate the largest uncertainty, there are still many steps required for a complete extraction of $\alpha_S$ from jet substructure.
The most important is to connect the topic fractions to the quark/gluon fractions in the theory calculations (recent progress in this direction can be found in Ref.~\cite{Stewart:2022ari}). There may also be different complications depending on which observables are used.
For example, the exclusive groomed jet mass cross section cleanly factorizes into quark and gluon components, but for the inclusive groomed jet mass and other observables like the energy-energy correlation function, the quark/gluon fraction is scale dependent~\cite{Frye:2016aiz,Frye:2016okc,Marzani:2017kqd,Marzani:2017mva,Kang:2018jwa,Kang:2018vgn,Chen:2020vvp}.
Beyond connecting topics with theory, further studies on the impact of hadronization and other non-perturbative effects will be required, with significant progress already underway~\cite{Hoang:2019ceu,Pathak:2020iue}.
To meet the high bar from the Particle Data Group~\cite{Zyla:2020zbs}, the precision on the perturbative side must also be improved.  One of the key ingredients is the $2\rightarrow 3$ jet cross section at next-to-next-to-leading order (NNLO), which was recently computed~\cite{Czakon:2021mjy} at leading color~\cite{Chen:2022tpk}.
Given the current tension in the global fit of $\alpha_S$, it is important to explore alternative measurements and jet substructure at the LHC may provide the necessary ingredients to build a deeper understanding of differences between the most precise methods.

\acknowledgments

We are grateful to Patrick~Komiske and Eric~Metodiev for collaboration during an early stage of this work, including the development of the codebase, and for feedback on the manuscript.
We thank Ian Moult, Jennifer Roloff and Jesse Thaler for insightful comments and helpful discussions, including on the manuscript, and Gregory Soyez for the (N)LL code for the differential cross section predictions in the resummation regime.  
This work was supported by the Office of Nuclear Physics of the U.S. Department of Energy (DOE) under grant DE-SC-0011090 and the DOE Office of High Energy Physics under grant DE-SC-0012567.
C. Sauer acknowledges support by the International Max Planck Research School for Precision Tests of Fundamental symmetries.

\bibliography{astopics}

\clearpage
\section*{Additional material}

\begin{figure}[htp]
\centering
\subfloat[Constituent multiplicity]{\includegraphics[width=0.99\columnwidth]{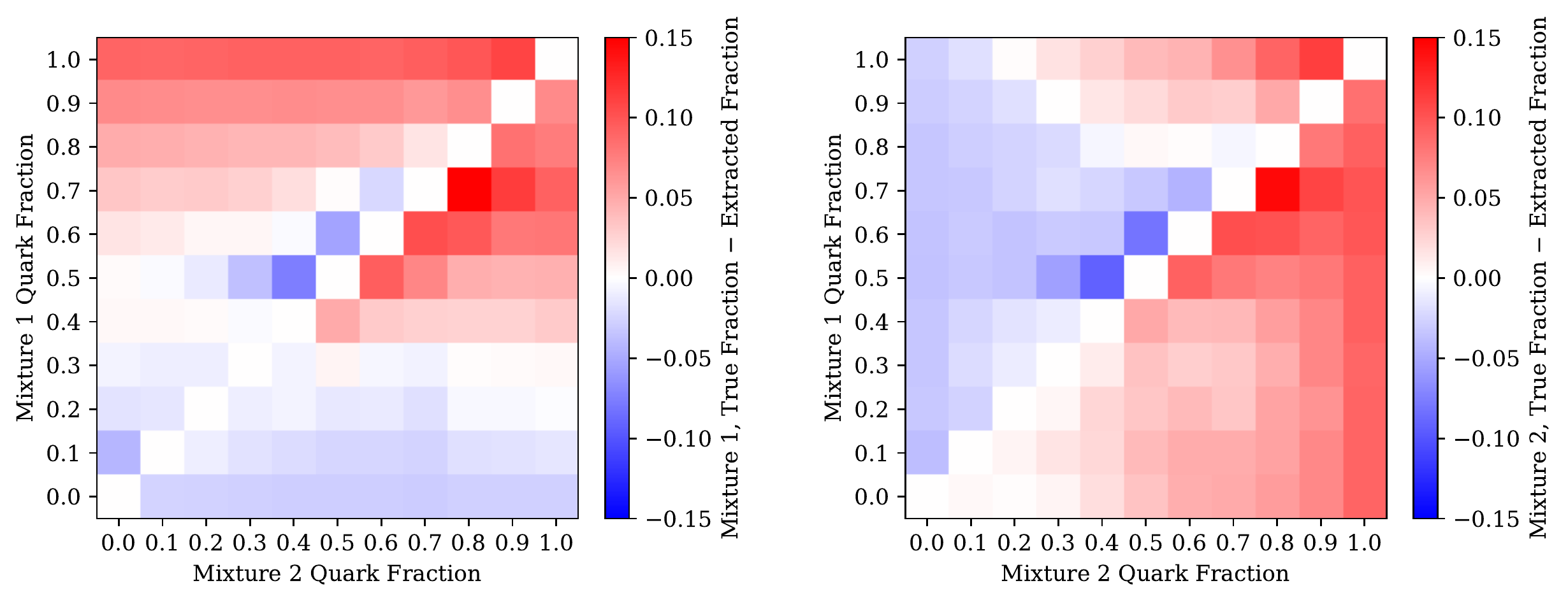}}\\
\subfloat[Particle flow network]{\includegraphics[width=0.99\columnwidth]{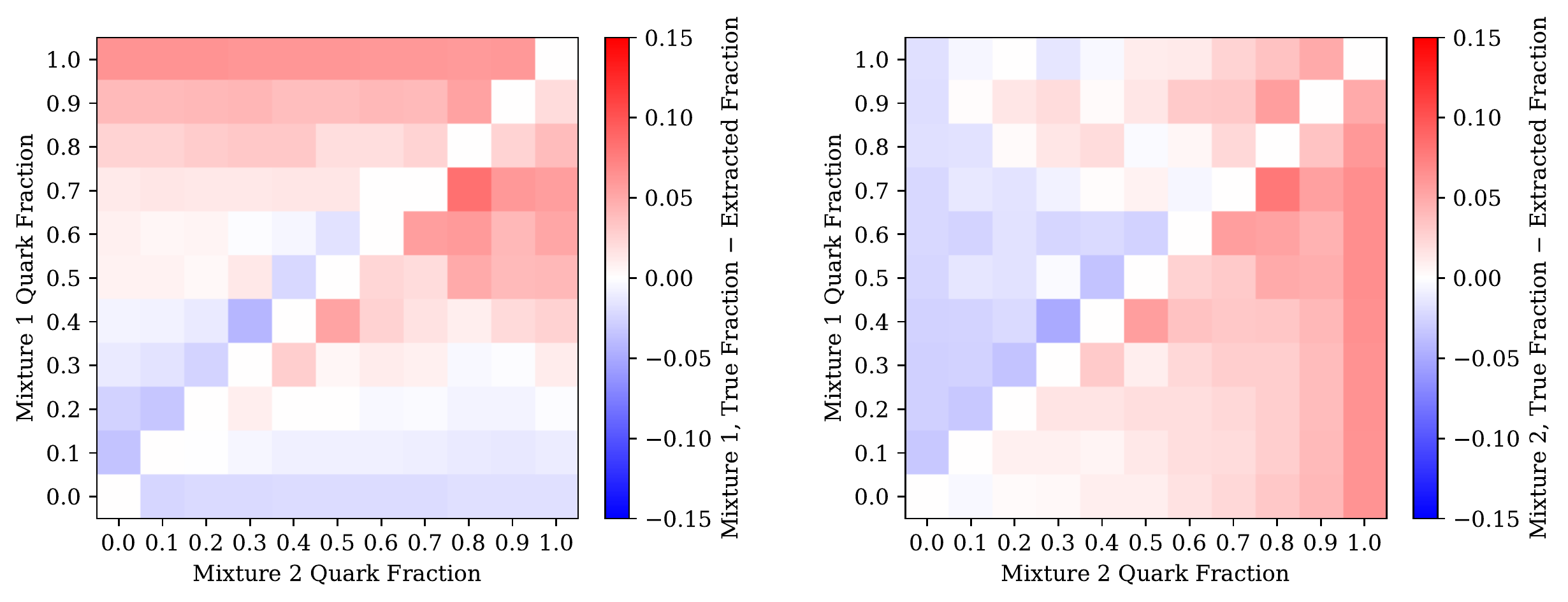}}
\caption{\label{fig:diffs}The difference between the extracted and true quark fractions of (left column) mixture 1 and (right column) mixture 2 for different quark fraction values between zero and one.
The (top row) jet constituent multiplicity or (bottom row) the output of a particle-flow network trained to discriminate quark and gluon jets was used to construct the topics used for the extraction of the quark fractions. 
The anchor regions used for the topics extractions as a function of this 2D plane are optimised for two mixed samples that are 50\% and 80\% quark jets, but could be reoptimized for each bin of the distribution in order to improve the closure for a specific set of mixtures.
The extracted fractions via PFN tend to exhibit smaller non-closure than those extracted using the particle multiplicity.
}
\end{figure}

\end{document}